\documentclass[runningheads]{llncs}
\usepackage{graphicx}
\usepackage{tabularx}

\usepackage{tikz}
\usepackage{times}  
\usepackage{helvet}  
\usepackage{courier}  
\usepackage[hyphens]{url}  
\usepackage{graphicx} 
\usepackage{caption} 
\usepackage{pgfplots}
\pgfplotsset{compat=1.17}

\begin{document}
\title{Two Heads are Better than One: A Bio-inspired Method for Improving Classification on EEG-ET Data}
\titlerunning{Two Heads are Better than One}
%
\makeatletter
\newcommand{\printfnsymbol}[1]{%
  \textsuperscript{\@fnsymbol{#1}}%
}
\makeatother

\author{Eric Modesitt\printfnsymbol{1}
    \and Ruiqi Yang\printfnsymbol{1}
    \and Qi Liu\thanks{These three authors contributed equally}
     }
\authorrunning{E. Modesitt et al.}
%
\institute{University of Illinois Urbana-Champaign\\
\email{ericjm4@illinois.edu}
\and
University of California Santa Barbara
\email{ruiqiyang@ucsb.edu}
\and
Palo Alto High School
\email{ql39535@pausd.us}}
\maketitle              
\begin{abstract}
Classifying EEG data is integral to the performance of Brain Computer Interfaces (BCI) and their applications. However, external noise often obstructs EEG data due to its biological nature and complex data collection process. Especially when dealing with classification tasks, standard EEG preprocessing approaches extract relevant events and features from the entire dataset. However, these approaches treat all relevant cognitive events equally and overlook the dynamic nature of the brain over time. In contrast, we are inspired by neuroscience studies to use a novel approach that integrates feature selection and time segmentation of EEG data. When tested on the EEGEyeNet dataset, our proposed method significantly increases the performance of Machine Learning classifiers while reducing their respective computational complexity. 

\keywords{Machine Learning \and SVM \and KNN \and Boosting \and Brain-Computer Interfaces \and EEG-ET Classification \and Feature Selection \and Bio-inspired Learning.}
\end{abstract}
\section{Introduction}
Electroencephalography (EEG) applications for classifying motor imagery tasks are significant for their medical purposes in assisting disabled civilians. However, historically Brain-Computer-Interfaces (BCI) have been limited by a lack of data, a low signal-to-noise ratio, and data's nonstationarity over time \cite{lotte2018review,lotte2007review,qu2020using}. 
 
\begin{figure} [b!]
    \centering
    \includegraphics[scale=.25]{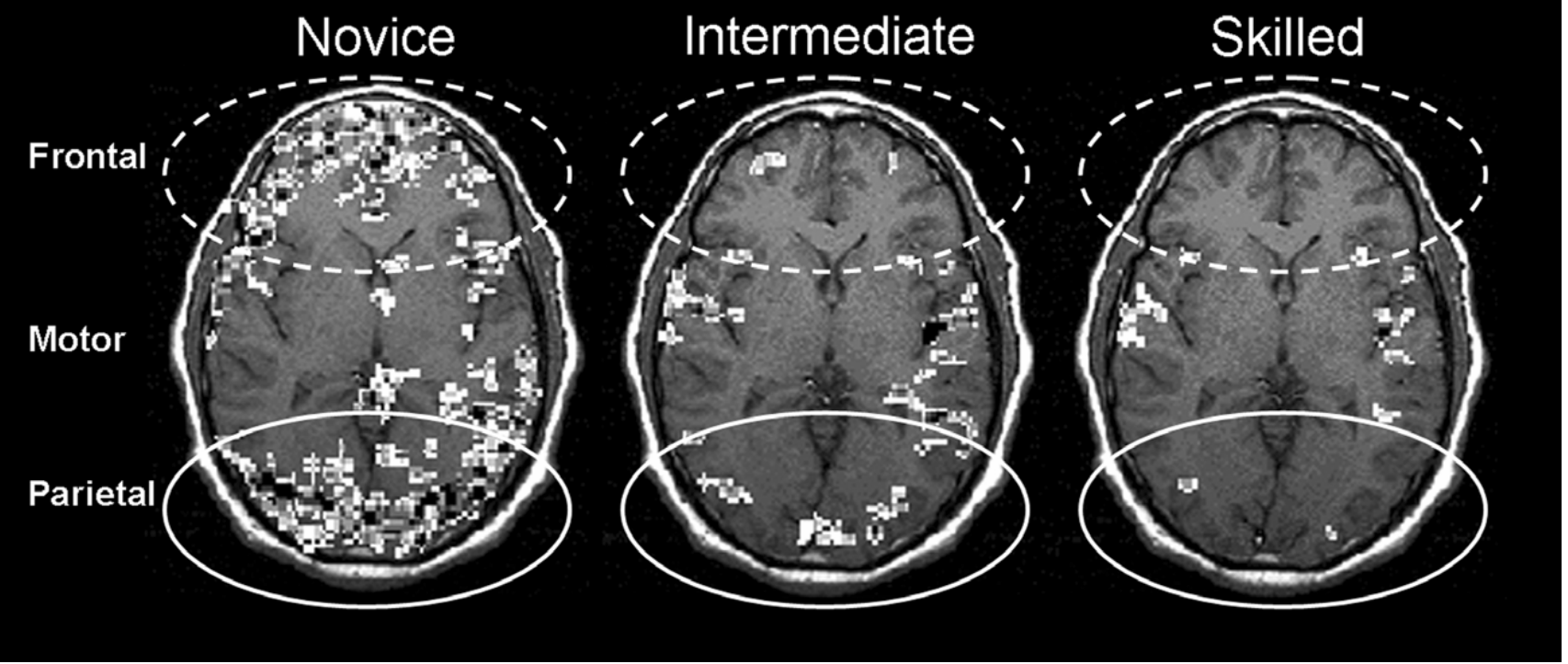}
    \caption{Neural Activity after 0 minutes (Novice), 30 minutes (Intermediate), and 60 minutes (Skilled) in a motor imagery task \cite{hill2006brain}}
    \label{fig:brain2}
\end{figure}

     Additionally, prior neuroscience studies show that patients' neural activity tends to continually decrease as a patient learns to complete a new task (Fig. \ref{fig:brain2}). Based on these results, we expect that different time-segmented partitions of EEG data produce vastly different feature representations. After all, according to previous neuroscience studies, well-learned tasks produce completely different neural signals than tasks that haven't been learned \cite{hill2006brain}, so it follows that distinct time-segmented data partitions should be treated as different. In this work, we were motivated by such findings to create the Two Heads method, where we split the EEG-ET (EEG-Eye Tracking) data by individual subject, concatenated the results, and applied feature selection. Our method was tested on EEGEyeNet's Left Right (LR) task dataset (publicly available) to classify saccade direction using EEG-ET data \cite{kastrati2021eegeyenet}.

\section{Related Work}

Classification models are typically designed in such a way that assumes every input corresponds with a fundamentally similar set of features to other inputs, thus allowing the models to learn these essential features \cite{deb2021trends,jiang2022deep,qu2020multi,tang2014data,zhao2021dsal}. Most popular feature selection methods for EEG take into account this typical design\cite{craik2019deep,qu2018eeg,roy2019deep}. Such approaches include simple statistical transformations such as Principal Component Analysis (PCA) and Independent Component Analysis (ICA) \cite{al2021deep,6696119,YU20141498}, the selection of filter banks \cite{PMID:29297303,ZHANG201585}, and other algorithms that transform high dimensional feature spaces based on individual model performance \cite{BAIG2017184,s17112576,TAN2020100597,lr_feature_selection}.

However, given the adaptive nature of the brain when learning a new task, it is unclear whether the features learned in one segment of our data would share underlying features with data in another time segment. Current approaches that work towards a solution to the non-stationarity of EEG include adding adaptiveness to the model in either feature extraction process \cite{li2017adaptive,mousavi2022spectrally,7038203} or the classification process \cite{abu2019,antony2022classification,lahane2019}. However, relevant solutions in the process of feature selection for EEG datasets are relatively lacking. Hence, we introduce the Two Heads method to address the dynamic nature of EEG data through a novel approach to selecting the best feature of subsets of data.

\section{Methods}
\subsection{Two Heads Data Segmentation}
\begin{figure} [b!]
    \centering
    \includegraphics[scale=.35]{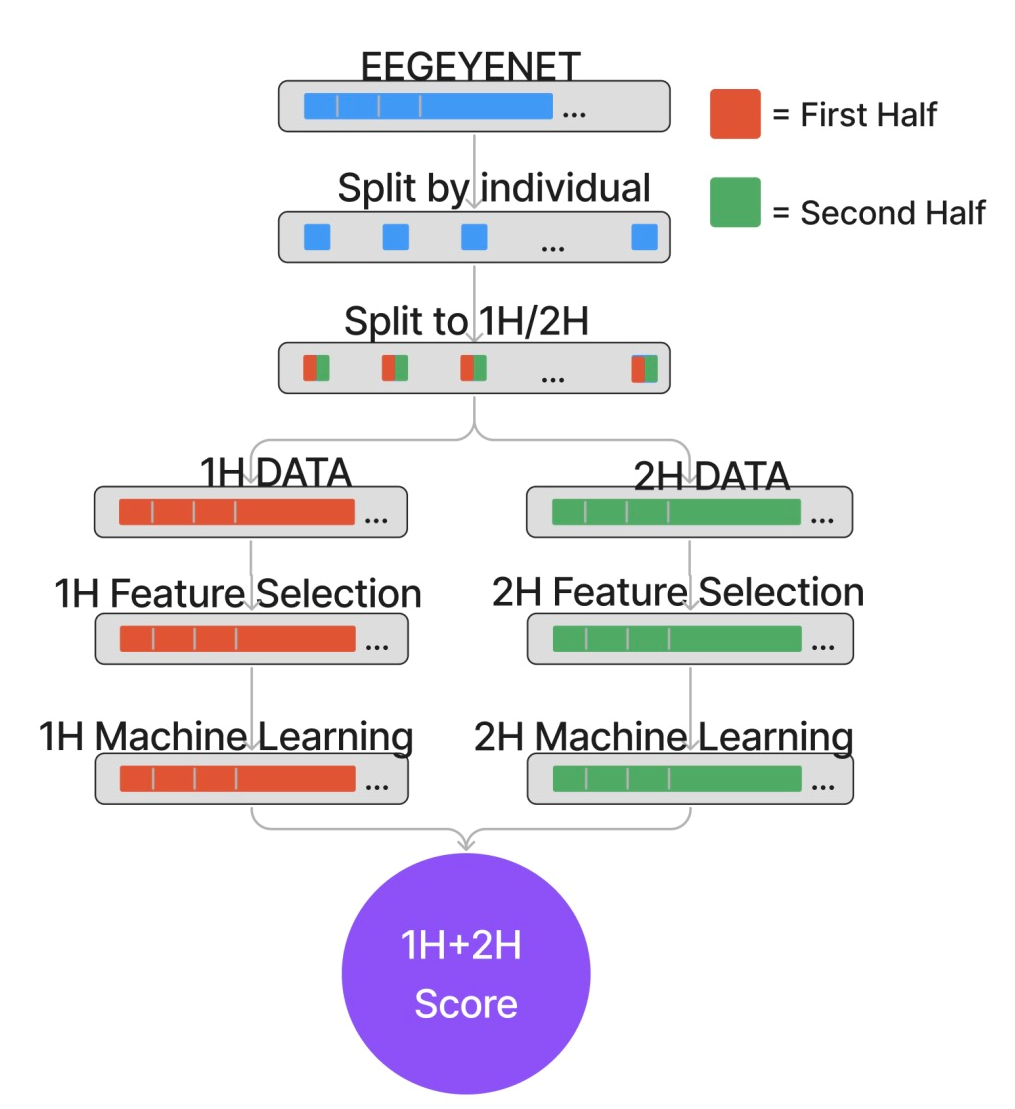}
    \caption{Model flowchart, from EEGEyeNet to splitting by individual, to taking 1H/2H, to separate feature engineering and classification}
    \label{fig:pipeline}
\end{figure}

 Previous work introduces this segmentation of learning as a three-stage process, where the learner progresses from Novice to Intermediate, and finally to Skilled (Fig. \ref{fig:brain2}). However, in this work, we decided to split the data into two parts. We made this deviation to reflect the difference in data collected per individual. Notably, the three fMRI shown in Figure 1 images were taken over a 60-minute interval per subject \cite{hill2006brain}, whereas the EEGEyeNet dataset only records over a  15-minute period per subject \cite{kastrati2021eegeyenet}. Therefore, to make up for the differences in length, we reduced the number of segmented stages of learning from three to two.

Starting from EEGEyeNet's baseline, we took EEG-ET data for each participant in the Left-Right task and modified the data. Accordingly, we split each individual subject's data into two disjoint time intervals of equal length. After splitting the first and second halves of individuals' data into groups 1H (1st Half) and 2H (2nd Half) respectively, we applied feature selection algorithms to these two separate data arrays.  

\subsection{Two Heads Feature Selection}
After our data segmentation process, we extracted features by following the existing pipeline suggested by EEGEyeNet benchmark. The feature extraction process includes band-pass filtering at the alpha band (8 - 13 Hz), and applying a Hilbert transform on the filtered data. Finally, the phase and amplitude for each of the 129 channels were extracted as features \cite{kastrati2021eegeyenet}. Then we applied feature selection to our two distinct groups of features, each representing different time intervals (1H and 2H), separately. There are three common ways to select features, filter-based feature selection, wrapper-based feature selection, and embedded methods \cite{khaire2022stability}. However, in this work, we only applied a filter-based feature selector due to concerns about the increased computational complexity that other feature selection methods could add. In particular, our results demonstrate our usage of Univariate Feature Selection from the implementation in the sklearn package, using the ANOVA F-value as our scoring mechanism. In the context of EEG, integrating these techniques could contribute to enhancing feature selection and classifier performance. (Figure \ref{fig:pipeline}) shows our entire pipeline.

\subsection{Two Heads Training}
As (Figure \ref{fig:pipeline}) shows, 1H and 2H feature sets were processed into separate training pipelines. The training pipeline split each segmented dataset into train, validation and test sets based on a proportion of 0.7:0.15:0.15. For our work, each of the eight ML classifiers shown in (Table \ref{table:accuracy}) was trained on five iterations of the same data, and we present the mean scores. Our results show little to no significant deviation in accuracy across iterations, resulting in us not reporting the standard deviation. The final score for each classifier is a weighted average of 1H and 2H classification accuracies. 

\section{Results}

\renewcommand{\arraystretch}{2.0}
\newcolumntype{P}[1]{>{\centering\arraybackslash}p{#1}}

\begin{table}[b!]
\begin{center}

\begin{tabular}{ |p{2.5cm}||P{1.5cm}|P{1.5cm}|P{1.7cm}|P{1.7cm}|P{1.5cm}| }

\hline
\textbf{ML Classifier}&\textbf{SOTA}&\textbf{FS}&\textbf{1H (1st Half)}&\textbf{2H (2nd Half)}& \textbf{1H+2H}\\
 \hline
 Gaussian NB & \centering87.7 & 90.8 & \centering\textbf{94.6} & 91.4 & 93.0\\
 
 LinearSVC & \centering 92.0 & 92.1 & \textbf{96.8} & 88.7 & 92.7\\
 
 KNN & \centering90.7 & 96.1 & \textbf{96.9} & 95.7 & 96.3\\

 RBF SVC/SVR & \centering89.4 & 96.5 & \textbf{97.5} & 95.9 & 96.7\\
  
 AdaBoost & \centering96.3 & 96.5 & \textbf{97.7} & 95.2 & 96.4\\

 Random Forest & \centering96.5 & 96.9 & \textbf{97.9} & 96.4 & 97.1\\
 
 Gradient Boost & \centering97.4 & 97.5 & \textbf{98.2} & 96.9 & 97.5\\
 
 XGBoost & \centering97.9 & 98.1 & \textbf{98.6} & 97.6 & 98.1\\[.25cm]
\hline

\end{tabular}
\end{center}
\caption{\textbf{Results (classification accuracies in \%)} for state-of-the-art (SOTA) benchmark from EEGEyeNet, traditional feature selection (FS) and our 1H+2H approach of implementing feature selection. Mean scores of 5 runs of each considered classifier.}
\label{table:accuracy}
\end{table}
(Table \ref{table:accuracy}) shows the mean scores from the SOTA benchmark in EEGEyeNet and the mean scores of ML classifiers coupled with feature selection models without segmenting the EEG data into different time series. As portrayed in the table, this traditional method of implementing feature selection has improved the performance of all ML classifiers from the original SOTA benchmark. 

The results demonstrated in (Table \ref{table:accuracy}) suggest that a traditional feature selection approach significantly improved all ML classifiers overall. However, the results of our Two Heads method in (Table \ref{table:accuracy}) suggest that our novel approach also improves the performance of ML classifiers on a  grander scale.


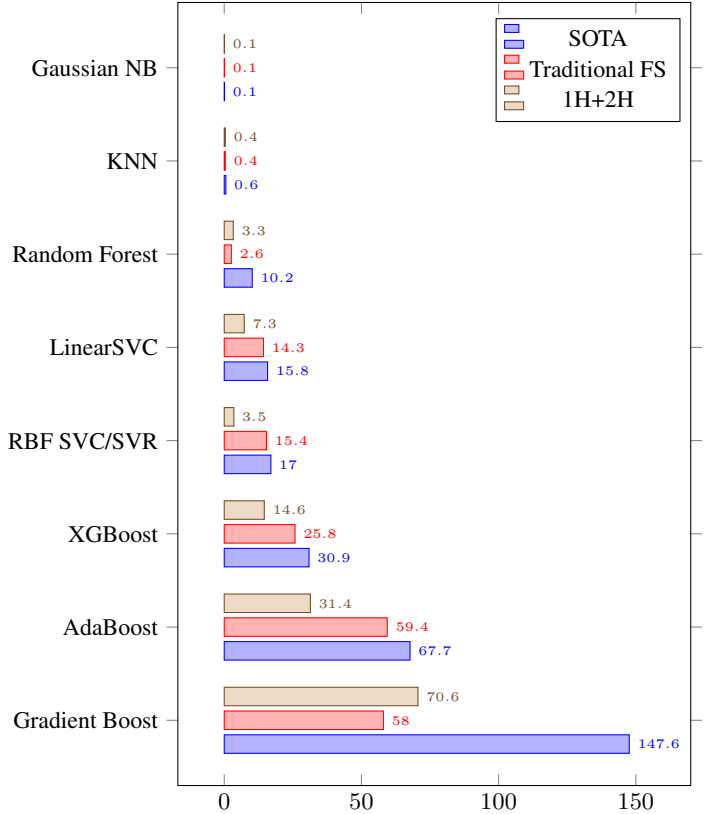
\begin{figure}[b!]
    \hskip-1.15cm
    \begin{tikzpicture}
    \centering
    \begin{axis} [xbar,xmax=170, height = 12cm, width = 8.4cm, symbolic y coords={Gradient Boost,AdaBoost,XGBoost,RBF SVC/SVR,LinearSVC,Random Forest,KNN,Gaussian NB}, bar width=7pt, nodes near coords,
        every node near coord/.append style={font=\tiny},]
        \addplot coordinates {
        (30.9,XGBoost)
        (147.6,Gradient Boost)
        (10.2,Random Forest)
        (67.7,AdaBoost)
        (17.0,RBF SVC/SVR)
        (0.6,KNN) 
        (15.8,LinearSVC) 
        (0.1,Gaussian NB) 

    };
    
    \addplot coordinates {
        (25.8,XGBoost)
        (58.0,Gradient Boost)
        (2.6,Random Forest)
        (59.4,AdaBoost)
        (15.4,RBF SVC/SVR)
        (0.4,KNN) 
        (14.3,LinearSVC) 
        (0.1,Gaussian NB)
    };
    
    \addplot coordinates {
        (14.6,XGBoost)
        (70.6,Gradient Boost)
        (3.3,Random Forest)
        (31.4,AdaBoost)
        (3.5,RBF SVC/SVR)
        (0.4,KNN)
        (7.3,LinearSVC)
        (0.1,Gaussian NB)
    };

    \legend { SOTA, Traditional FS, 1H+2H}
    \end{axis}
    
    \end{tikzpicture}
    \caption{\textbf{Comparison of Run Time (Seconds)} for state-of-the-art (SOTA) benchmark from EEGEyeNet, traditional feature selection (FS) method, and our 1H+2H approach of implementing feature selection. The mean of 5 trials for each method is measured.}
    \label{fig:runtime}
\end{figure}

In addition to our novel approach's contribution to expanding the marginal increase of ML classifiers' performance in comparison to both the SOTA benchmark and traditional feature selection, our procedure also manages to achieve a lower or equal runtime on all ML classifiers tested except random forest and gradient boost, as shown in (Figure \ref{fig:runtime}). It suggests that our approach is promising not only in achieving higher accuracy but also in reducing computation complexity -- which optimizes the performance-to-computation ratio for ML classifiers in comparison with current procedures.

Moreover, the result for 1H based on XGBoost even exceeds some deep learning classifiers. As tested in the EEGEyeNet benchmark, CNN achieves a mean accuracy of 98.3 \cite{kastrati2021eegeyenet} which is lower than the 98.6 mean score by our 1H approach using XGBoost.

\section{Discussion}
The comparison of our results to those from EEGEyeNet's SOTA benchmark and the traditional feature selection method suggests that our approach increases the performance of ML classifiers by a  significant margin. In addition, although general feature selection also decreases the runtime for ML models to finish training, our novel method achieves an even lower runtime involved in training most of the ML models. Generalizing these accurate results suggests a new method for researchers to maximize the accuracy of ML classifiers while minimizing their computation complexity (namely, feature selection and splitting data into different time intervals). This could potentially lower the barrier to entry for EEG classification tasks, as successful future implementations of this method should guarantee that people would need fewer data samples and thus less computational power to achieve cutting-edge results. 

Our work has two main directions for future work. Currently, our work has been tested only on the EEGEyeNet dataset. We would need more evidence to claim with certainty that this result generalizes to other EEG datasets such as GigaDB \cite{cho2017eeg} and BCI Competition IV dataset \cite{tangermann2012review}. Second, we tested our results on several machine learning models. We can not generalize that all types of predictive models (especially deep learning models) will follow the same trend of increased performance given our Two Heads method. Future work will focus on testing our method in SOTA deep learning architectures, including CNN, AutoEncoders \cite{al2021deep}, and so on. Other machine learning analysis for biomedical data \cite{al2021deep,bashivan2015learning,chedid2022development,deb2022systematic,qian2021two,qu2020identifying,saeidi2021neural,yi2022attention} or other time series data \cite{chen2021data,lu2022cot,luo2022multisource,ma2020statistical,vaswani2017attention,zhang2021trust,zhangtrep,zhang2022attention} could be worth trying to compare with our approach on the same dataset. We would also like to examine the robustness of the models we have investigated, given the widespread attention to the vulnerability of machine learning models \cite{jiang2022camouflaged}.

Finally, we urge future research to take into account the dynamic nature of the brain in their experiments. For example, researchers might create an ML model which helps steer a cursor on a screen for paralyzed patients. In this case, researchers could strategically focus their models on the segment of data where patients have learned how to complete the task, as this subset of data would produce a distinct result from the data collected while the patient is still learning. To tackle this issue, transfer learning can be employed to design a task-specific model with minimum effort of fine-tuning~\cite{an2020transfer}. Regardless, this result motivates us to treat EEG data collected on novel tasks differently from EEG data collected on more well-learned tasks.
        
\section{Conclusion} 
This paper proposes and tests a novel bio-inspired method to implement feature selection with ML classifiers. Through this approach, we found that the Two Heads method optimizes ML classifiers' performance-to-computation ratio and outperforms both the SOTA benchmark on EEGEyeNet and the traditional feature selection approach. Our findings suggest that time segmentation has a notable impact on EEG signals and prompts future research to consider the time component of EEG data during the data collection process and the training of ML classifiers.
  
%
%
%
\bibliographystyle{splncs04}

\bibliography{HCII23}

\end{document}